\documentclass[preprint,showpacs,preprintnumbers,amsmath,amssymb]{revtex4}

\usepackage{graphicx}
\usepackage{dcolumn}
\usepackage{bm}
\usepackage{color}
\usepackage{hyperref}
\usepackage{amssymb}
\usepackage{amsmath,empheq}
\usepackage{slashed}

\usepackage{tikz}
\usepackage{bm,bbm}
\usepackage{xcolor}
\usepackage{physics}
\usepackage{caption}
\usepackage{subcaption}

\newcommand{\be}{\begin{eqnarray}}
\newcommand{\ee}{\end{eqnarray}}
\newcommand{\id}{\mathbbm{1}}
\makeatletter
\def\l@subsection#1#2{}
\def\l@subsubsection#1#2{}
\makeatother
\numberwithin{equation}{section}

\begin{document}

\title{Boundary Condition Analysis of \\ First and Second Order Topological Insulators}

	\author{Xi Wu}
	\email{wuxi5949@gmail.com}
	\affiliation{School of Physics and Electronics, Hunan University, Changsha 410082, China}
	\author{Taro Kimura}
	\email{taro.kimura@u-bourgogne.fr}
	\affiliation{Institut de Mathématiques de Bourgogne, Universitée Bourgogne Franche-Comté, Dijon, France}

\begin{abstract}
We analytically study boundary conditions of the Dirac fermion models on a lattice, which describe the first and second order topological insulators.
We obtain the dispersion relations of the edge and hinge states by solving these boundary conditions, and clarify that the Hamiltonian symmetry may provide a constraint on the boundary condition.
We also demonstrate the edge-hinge analog of the bulk-edge correspondence, in which the nontrivial topology of the gapped edge state ensures gaplessness of the hinge state.
\end{abstract}

\maketitle

\tableofcontents

\section{Introduction}
In quantum mechanics, physical observables are obtained as the eigenvalue of Hermitian operators.
In the bulk system, the momentum operator $\hat{k} = - i \partial$ is a well-defined Hermitian operator as we have the translation symmetry.
However, if there is a boundary, the translation symmetry is partly violated, so that we have to be careful of the Hermiticity of the momentum operator.
In fact, in order to show the Hermiticity of the momentum operator, we shall use the integration by parts, which in principle could have the boundary contribution.


The existence of the boundary is essential in the study of topological materials, which exhibit nontrivial degrees of freedom localized on the boundary.
It is known that for topological materials, the band topology is characterized by certain points (called Dirac points) in the Brillouin zone~\cite{Hasan:2010to,Qi:2010qag}.
Hence, it is important to study the effective Hamiltonian, which describes the behavior in the vicinity of these points. 
From this point of view, it has been shown in \cite{Witten:2015aoa} that, in the continuum limit, the Hermitian property of the Hamiltonian demands boundary conditions for the Dirac fermion system.

In the context of microscopic models of the topological materials defined on a lattice, we should consider several combinations of the gamma matrices, e.g., $\Gamma_i \cos k+\Gamma_j \sin k$, to gain nontrivial topological properties in the Wilson fermion model~\cite{Wilson_1977}, the Su--Schrieffer--Heeger (SSH) mode~\cite{Su:1979ut} and Haldane's model (Chern insulator)~\cite{Haldane:1988uf}, and so on.
In the presence of such a combination term, analysis of the boundary condition becomes in fact involved compared to the ordinary Dirac fermion system.
See, for example,~\cite{Isaev:2011gq,Enaldiev_2015,Hashimoto:2016kxm,Kimura2018,Fukui:2020wo} for the related works on the boundary conditions of the topological materials.
%
%
%
The purpose of this paper is to explore the boundary condition obtained from the Hermitian property of the lattice model involving several combinations of the gamma matrices.
We in particular show how the boundary condition affects the physical properties of the edge state localized on the boundary, including the energy spectrum and the penetration depth.
Moreover, we consider an intersection of two different boundaries.
Imposing the compatibility of the boundary conditions, we may obtain the localized state at the intersection of the boundaries, which is a key feature of the higher-order topological insulators~\cite{Benalcazar_2017,Hayashi:2016dec,Hashimoto:2017ta,Benalcazar:2017dhp,Song:2017uhz,Schindler2018,Langbehn_2017}.




This paper is organized as follows: in Sec.~\ref{BCL}, we use the Hermiticity property of the Dirac Hamiltonian to derive boundary conditions for the edge and bulk states; in Sec.~\ref{1OTI}, we study edge state wave functions and dispersion relations in two examples of the first order topological insulators from the point of view of the boundary conditions; in Sec.~\ref{2OTI} we study wave functions and dispersion relation of the edge and hinge states for a model of the second-order topological insulator; in Sec.~\ref{sum} we conclude with a summary and discussion.
\section{Boundary conditions on the lattice}
\label{BCL}
In this section, we derive primitive lattice boundary conditions in a general setup based on the Hermiticity of the difference operator.
We show that the boundary conditions are implemented for edge states in a simple form, which is analogous with the continuum model, while a slight different treatment is necessary for the bulk states.

In order to obtain a topologically nontrivial phase, we should include the momentum-dependent mass term in the Dirac Hamiltonians on a lattice, that is known as the Wilson term.
Let us consider the following one-dimensional tight-binding model:
\be\label{DH1}
    \mathcal{H}_\text{1d} =
	\sum_{n=1}^N\psi^{\dagger}_n(\sigma_1\cos \hat{k}+\sigma_2\sin\hat{k})\psi_n=\sum_{n=1}^N\frac{1}{2}\psi^{\dagger}_n(\sigma_1( \nabla+ \nabla^{\dagger}+2)-i\sigma_2( \nabla- \nabla^{\dagger}))\psi_n\,.
\ee
where $\sigma_{1,2}$ are the Pauli matrices, and we define the difference operator, 
\begin{subequations}\label{nas}
 \begin{align}\label{na}
 \nabla \psi_n & := \psi_{n+1} - \psi_n=(e^{i\hat{k}}-1) \psi_n
 \, , \\
 \nabla^{\dagger} \psi_n & := \psi_{n-1} - \psi_n=(e^{-i\hat{k}}-1) \psi_n
 \, ,\label{nad}
\end{align}
\end{subequations}
with the momentum operator $\hat{k} = - i \partial$.
Requiring the Hermiticity of the Hamiltonian, Eq.~(\ref{DH1}) should be also written as
\be
	\sum_{n=1}^N\psi^{\dagger}_n(\sigma_1\cos \hat{k}+\sigma_2\sin\hat{k})\psi_n = 
	\sum_{n=1}^N((\sigma_1\cos \hat{k}+\sigma_2\sin\hat{k})\psi_n)^{\dagger}\psi_n\,.
	\label{DH1_eq}
\ee
In order to obtain this equality, we shall impose the boundary condition as follows.

\subsection{Derivation of the boundary condition}

By definition of the difference operator~\eqref{nas}, we first obtain
\begin{subequations}
\be
	\psi^{\dagger}_n\nabla\psi_n 
	& = \psi^{\dagger}_n\psi_{n+1}-\psi^{\dagger}_{n-1}\psi_{n}+(\nabla^{\dagger}\psi_n)^{\dagger}\psi_n \,,\\
	\psi^{\dagger}_n\nabla^{\dagger}\psi_n 
	& = \psi^{\dagger}_n\psi_{n-1}-\psi^{\dagger}_{n+1}\psi_{n}+(\nabla\psi_n)^{\dagger}\psi_n\,,
\ee
\end{subequations}
where we use the following relations,
\begin{subequations}
\be
	(\nabla^{\dagger}\psi_n)^{\dagger}\psi_n
	& = (\psi^{\dagger}_{n-1}-\psi^{\dagger}_n)\psi_n \, , \\
	(\nabla\psi_n)^{\dagger}\psi_n
	& = (\psi^{\dagger}_{n-1}-\psi^{\dagger}_n)\psi_n \,.
\ee
\end{subequations}
Summing over the site, we then obtain the relations,
\begin{subequations}\label{pnp}
\begin{align}
	\sum_{n=1}^N\psi^{\dagger}_n\nabla\psi_n & = \psi^{\dagger}_N\psi_{N+1}-\psi^{\dagger}_{0}\psi_{1}+\sum_{n=1}^N(\nabla^{\dagger}\psi_n)^{\dagger}\psi_n\,, \\
	\sum_{n=1}^N\psi^{\dagger}_n\nabla^{\dagger}\psi_n & = \psi^{\dagger}_{1}\psi_{0}-\psi^{\dagger}_{N+1}\psi_{N}+\sum_{n=1}^N(\nabla\psi_n)^{\dagger}\psi_n\,.
\end{align}
\end{subequations}
These relations are interpreted as a difference analog of integration by parts, which provides an extra contribution if there exists the boundary,
\begin{align}
    \int \dd{x} \psi^\dag \partial \psi = (\psi^\dag \psi)\Big|_\text{boundary} -\int \dd{x} (\partial \psi^\dag) \psi  
    \, .
\end{align}

We consider the following combinations that appear in the Hamiltonian \eqref{DH1}.
From the integration by parts relations \eqref{pnp}, we obtain
\begin{subequations}\label{cos_sin}
\be\label{cos}
	\sum_{n=1}^N\psi^{\dagger}_n\cos \hat{k}\,\psi_n 
	&=& \sum_{n=1}^N \psi^{\dagger}_n \qty( \frac{\nabla+ \nabla^{\dagger}}{2}+1)\psi_n\nonumber \\ 
	&=& \sum_{n=1}^N(\cos \hat{k}\,\psi_n)^{\dagger}\psi_n+\frac{1}{2}(\psi^{\dagger}_N\psi_{N+1}-\psi^{\dagger}_{N+1}\psi_{N}-\psi^{\dagger}_{0}\psi_{1}+\psi^{\dagger}_{1}\psi_{0})
	\,, \\
	\sum_{n=1}^N\psi^{\dagger}_n\sin \hat{k}\,\psi_n &=& \sum_{n=1}^N\psi^{\dagger}_n \qty( \frac{\nabla- \nabla^{\dagger}}{2i} ) \psi_n\nonumber \\
	&=& \sum_{n=1}^N(\sin \hat{k}\,\psi_n)^{\dagger}\psi_n+\frac{1}{2i}(\psi^{\dagger}_N\psi_{N+1}+\psi^{\dagger}_{N+1}\psi_{N}-\psi^{\dagger}_{0}\psi_{1}-\psi^{\dagger}_{1}\psi_{0})
	\,.
\ee
\end{subequations}
Hence, the Hamiltonian \eqref{DH1} is written as follows,
\be\label{cossin}
	\sum_{n=1}^N\psi^{\dagger}_n(\sigma_1\cos \hat{k}+\sigma_2\sin\hat{k})\psi_n &=& \sum_{n=1}^N((\sigma_1\cos \hat{k}+\sigma_2\sin\hat{k})\psi_n)^{\dagger}\psi_n
	\nonumber \\
	&& +\frac{1}{2}(\psi^{\dagger}_N\sigma_1\psi_{N+1}-\psi^{\dagger}_{N+1}\sigma_1\psi_{N}-\psi^{\dagger}_{0}\sigma_1\psi_{1}+\psi^{\dagger}_{1}\sigma_1\psi_{0})\nonumber
	\\&&+\frac{1}{2i}(\psi^{\dagger}_N\sigma_2\psi_{N+1}+\psi^{\dagger}_{N+1}\sigma_2\psi_{N}-\psi^{\dagger}_{0}\sigma_2\psi_{1}-\psi^{\dagger}_{1}\sigma_2\psi_{0})\,.
\ee
In order that the equality~\eqref{DH1_eq} holds, the boundary terms in Eq.~(\ref{cossin}) should vanish.
This imposes the boundary condition.

\subsection{Analysis of the boundary condition}

There are two possibilities for the boundary condition.
The first is the periodic boundary condition,
\be
	\psi_n=\psi_{n+N} \, \qquad \forall n \in \{1, \ldots, N\}
\ee
and the other is the open boundary condition,
\begin{subequations}
 \begin{empheq}[left=\empheqlbrace]{align}
		&\label{bc01} \psi^{\dagger}_{0}\sigma_1\psi_{1}-\psi^{\dagger}_{1}\sigma_1\psi_{0}-i(\psi^{\dagger}_{0}\sigma_2\psi_{1}+\psi^{\dagger}_{1}\sigma_2\psi_{0})=0 \\ 
		\label{bcNN1} &\psi^{\dagger}_N\sigma_1\psi_{N+1}-\psi^{\dagger}_{N+1}\sigma_1\psi_{N}+i(\psi^{\dagger}_N\sigma_2\psi_{N+1}+\psi^{\dagger}_{N+1}\sigma_2\psi_{N})=0,
\end{empheq}
\end{subequations}
where the two boundary contributions vanish independently. 
Since these two equations have similar structure, we focus on the first equation (\ref{bc01}). 
Noticing the relation $\psi_1=e^{i\hat{k}}\psi_0$, we may write the boundary condition (\ref{bc01}) locally. 
We discuss the bulk and the edge states separately in the following.

\subsubsection{Edge state}

For an edge state localized on the boundary, we assume that the wave function takes the following form $\psi_n =\beta \psi_{n-1}$ where $\beta \in \mathbb{R}$. 
We also impose the normalizability condition $|\beta| < 1$.
Then, from the boundary condition (\ref{bc01}), we obtain
\be\label{ebc}
	\psi_0^{\dagger}\sigma_2\psi_0=0 \, . 
\ee
In fact, the $\sigma_1$-term does not play a role in the boundary condition for the edge state.
Notice that this result is straightforwardly generalized to arbitrary dimensions.
In general, the boundary condition \eqref{ebc} is interpreted as the no in/out-going current condition~\cite{Kimura2018}.

\subsubsection{Bulk state}

For a bulk state, we take the Fourier transform, and the differential operator $\hat{k}$ may be replaced with the corresponding real eigenvalue $k$.
Noticing
\be
	\psi_1 =e^{i{k}} \psi_{0}\,,	~~ \psi_1^{\dagger} =e^{-i{k}} \psi_{0}^{\dagger}\,,
\ee
and from the boundary condition (\ref{bc01}), we have
\be \label{bubc}
	\psi_0^{\dagger}(\sigma_1\sin {k}-\sigma_2\cos {k} )\psi_0=0\,.
\ee
Namely, the boundary condition depends on momentum $k$ in general.
We remark that in the limit $k \to 0$, Eq. (\ref{bubc}) reduces to Eq. (\ref{ebc}). 

\section{First order topological insulator and edge states under lattice boundary conditions}
\label{1OTI}
In this section we discuss the dispersion relation for generic edge states based on the boundary conditions for one-dimensional Su--Schrieffer--Heeger (SSH) model and two-dimensional Wilson fermion model (Chern insulator).
As we consider the boundary condition in one direction, this situation corresponds to the first order topological insulator.
Moreover, in the case of one-dimensional SSH model, although the gapless edge state is protected by chiral symmetry, the boundary condition in general violates it, and thus the edge state is gapped out; In the case of two-dimensional Wilson fermion model, the chiral gapless state is topologically protected under variation of boundary conditions. 

\subsection{One-dimensional SSH model}

The Hamiltonian of the one-dimensional SSH model can be written as
\be
    \mathcal{H}_\text{SSH} = \sum_{n=1}^N \psi_n^\dag H_\text{SSH}(\hat{k}) \psi_n
    \, , \qquad
	H_\text{SSH}(\hat{k}) = (s+t\cos \hat{k})\sigma_1+ t\sin \hat{k}\sigma_2\,.
\ee
This Hamiltonian has the chiral symmetry, $\{H_\text{SSH}(\hat{k}), \sigma_3 \} = 0$, so that it is classified into the class AIII system.
Considering the edge states localized at $n=1$, we have the boundary condition
 \be
 	\psi_1^{\dagger}\sigma_2 \psi_1=0\,,
 \ee 
 with a generic solution 
 \be \label{bcssh}
 	\psi_1 = \left(\begin{array}{c}\cos \theta \\\sin \theta \end{array}\right)
 	\, ,
 \ee
where $\theta \in [0,2\pi)$ is a periodic parameter characterizing the boundary condition. 
We may apply the formalism discussed in our previous papers to explore this situation~\cite{Hashimoto:2016kxm,Hashimoto:2017ta}.
We assume that the edge state wave function takes the form of
\be
	\psi_n=\beta^{n-1} \psi_1
	\label{eq:edge_wf}
\ee
with $\beta \in \mathbb{R}$ and $|\beta| < 1$. 
Recalling that 
\be
	\cos \hat{k}=\frac{1}{2}( \nabla+ \nabla^{\dagger}+2)
	\, , \qquad 
	\sin \hat{k}=\frac{1}{2i}( \nabla- \nabla^{\dagger})
	\, ,
\ee
these terms may be replaced as follows for the edge state,
\be
	\cos \hat{k} \ \longrightarrow \ \frac{1}{2}\qty( \beta+\frac{1}{\beta})
	\, , \qquad
	\sin \hat{k} \ \longrightarrow \ \frac{1}{2i}\qty( \beta-\frac{1}{\beta})\,.
\ee
Then, the eigenvalue equation $(h_\text{SSH}(\hat{k})-\epsilon)\psi_n=0$
can be written as
\be\label{HeWbc}
	\left(\begin{array}{cc}-\epsilon & s+t\beta^{-1} \\ s+t\beta  & -\epsilon\end{array}\right)\beta^{n-1}\left(\begin{array}{c}\cos \theta \\\sin \theta \end{array}\right)=0\,, 
\ee
which gives rise to
\begin{subequations}
 \begin{empheq}[left=\empheqlbrace]{align}
		&\epsilon=\tan \theta (s+t\beta^{-	1})=\cot \theta (s+t\beta) \label{dispssh}\, ,
		\\  
		&
		(s+t\beta^{-1})\cos^2 \theta-(s+t\beta)\sin^2 \theta=0
		\label{betassh}\, .
\end{empheq}
\end{subequations}

Next we discuss the violation of symmetries by the boundary conditions. 
The boundary condition can be rewritten as follows,
\be \label{bcssh2}
	(1 - \sigma_1 \sin 2\theta - \sigma_3 \cos 2\theta )\psi_1=0
\ee
as in the matrix form, we have
\be
	0&=&\left(\begin{array}{cc}1-\cos 2\theta & -\sin 2\theta \\ -\sin 2\theta & 1+\cos 2\theta\end{array}\right)\psi_1 \nonumber \\
	&=&\left(\begin{array}{cc} 2 \sin \theta & 0 \\ 0 & -2\cos \theta\end{array}\right)\left(\begin{array}{cc}\sin \theta & -\cos \theta \\ \sin \theta & -\cos \theta \end{array}\right)\psi_1\,.
\ee
Now the boundary condition~(\ref{bcssh2}) is not compatible with the chiral symmetry of the original Hamiltonian, $\psi \to \sigma_3 \psi$, unless $\sin 2\theta=0$. 
Meanwhile from the dispersion Eq.~(\ref{dispssh}), we can see that the edge state has a non-zero energy unless $\beta=-s/t$ or  $\beta^{-1}=-s/t$, which correspond to $\cos \theta=0$ or $\sin \theta=0$ $(\iff \sin 2 \theta = 0)$ as seen from Eq.~(\ref{betassh}). 
Therefore, the edge state is gapless (zero-energy state) if the chiral symmetry is preserved, while it would be gapped (non-zero-energy state) if the chiral symmetry is violated due to the boundary condition.

\subsection{Wilson fermion}

Let us consider the Wilson fermion model in two dimensions:
\be
	H_\text{W}(\hat{k})= \sigma_1 (\cos \hat{k}_1+\cos \hat{k}_2-m-2) + \sigma_2 \sin \hat{k}_2 + \sigma_3 \sin \hat{k}_1 \,.
\ee
There is no specific symmetry for this model, so that it is classified into the class A system.
We assign the boundary condition at the boundary $n_2 = 1$, and we keep the $n_1$-direction as a bulk direction. 
Hence, we take the Fourier transform only for the $n_1$-direction to consider the wave function $\psi_{n_2}(k_1)$.

Now the boundary condition is given as follows,
 \be
 	\psi_1^{\dagger} (k_1) \sigma_2 \psi_1 (k_1) =0\,,
 \ee 
which is formally the same as the SSH model.
Hence, we have the same solution~\eqref{bcssh} with the parameter $\theta$ as before, and the wave function is given as
\be
	\psi_n(k_1)=\beta^{n-1} \psi_1(k_1)
\ee
with $\beta \in \mathbb{R}$ and $|\beta| < 1$.
In this case, we may replace
\begin{gather}
    \hat{k}_1 \ \longrightarrow k_1
    \, , \qquad
    \cos \hat{k}_2 \ \longrightarrow \ \frac{1}{2} \qty( \beta + \frac{1}{\beta} )
    \, , \qquad
    \sin \hat{k}_2 \ \longrightarrow \ \frac{1}{2i} \qty( \beta - \frac{1}{\beta} )
    \, .
\end{gather}
Then, the eigenvalue equation $(H_\text{W}(\hat{k})-\epsilon)\psi_n(k)=0$
can be written as follows,
\be\label{HeWbc}
	\left(\begin{array}{cc}\sin k_1-\epsilon & \cos k_1-m-2+\beta^{-1} \\ \cos k_1-m-2+\beta  & \sin k_1-\epsilon\end{array}\right)\beta^{n-1}\left(\begin{array}{c}\cos \theta \\\sin \theta \end{array}\right)=0\,,
\ee
together with
\be
	\det \left(\begin{array}{cc}\sin k_1-\epsilon & \cos k_1-m-2+\beta^{-1} \\ \cos k_1-m-2+\beta  & \sin k_1-\epsilon\end{array}\right)=0
	\, .
\ee

From the eigenvalue equation (\ref{HeWbc}) we can determine the parameter $\beta$ and the energy eigenvalue $\epsilon$ as follows:
Denoting $A:= -\sin k_1 \sin 2\theta +(\cos k_1-m-2) \cos2\theta$, we obtain a quadratic equation for the parameter $\beta$,
\be \label{beta}
	\beta^2 \cos^2 \theta +A\beta -\sin^2 \theta=0
\ee
which is solved by
\be \label{bepm}
	\beta_{\pm} =\frac{-A\pm\sqrt{A^2+\sin^2 2\theta}}{2\cos^2 \theta}\,.
\ee
 \begin{figure}[th]
 \begin{center}
    \begin{subfigure}[b]{0.3\textwidth}
         \includegraphics[width=\textwidth]{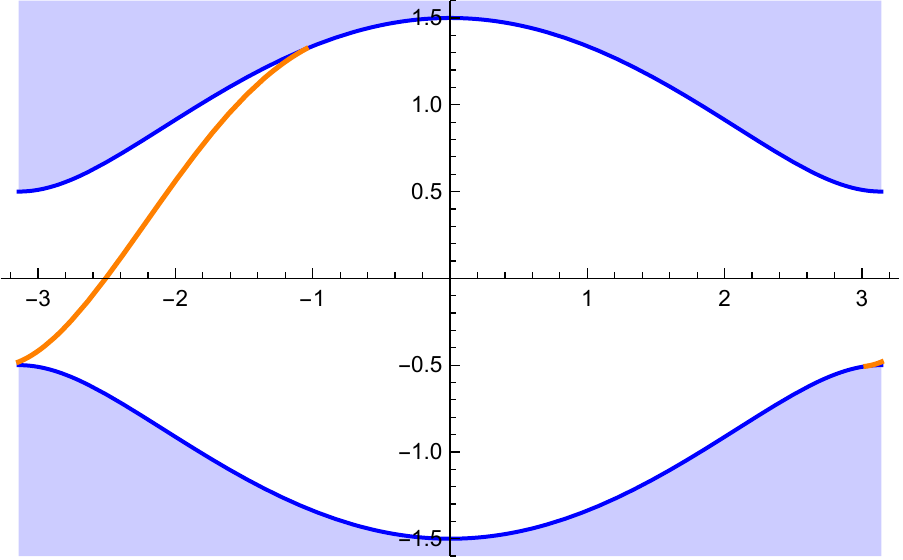}
         \caption{$\theta = 2\pi/7$}
         \label{fig:d1}
     \end{subfigure}
    \hfill
    \begin{subfigure}[b]{0.3\textwidth}
         \includegraphics[width=\textwidth]{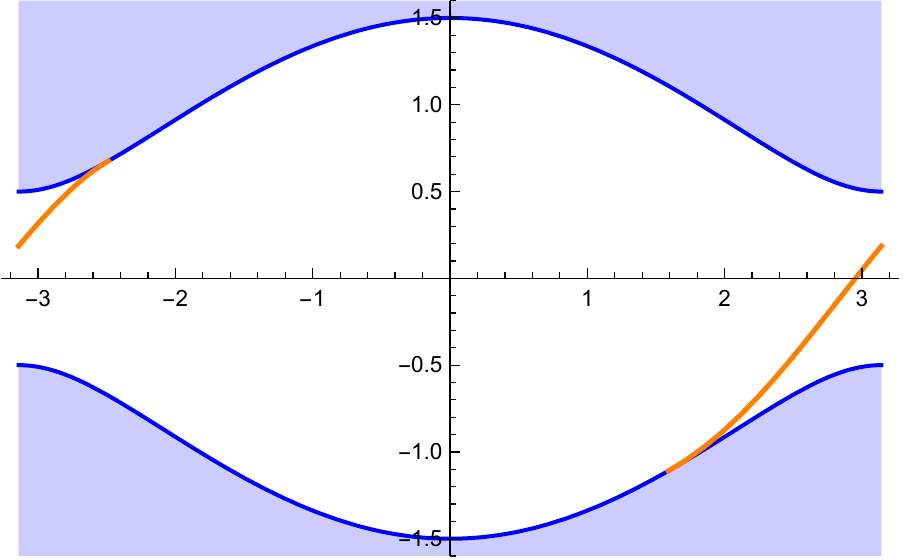}
         \caption{$\theta = 4\pi/7$}
         \label{fig:d2}
     \end{subfigure}
    \hfill    
    \begin{subfigure}[b]{0.3\textwidth}
         \includegraphics[width=\textwidth]{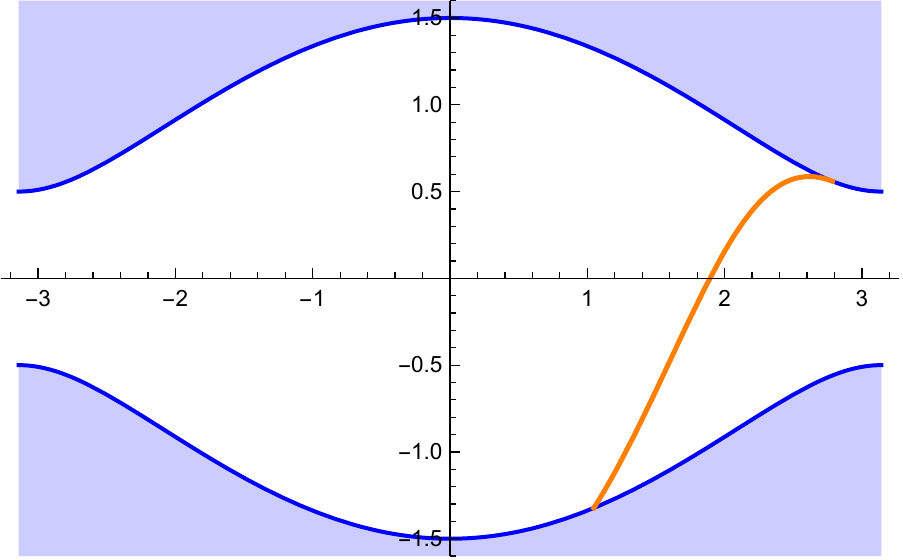}
         \caption{$\theta = 6\pi/7$}
         \label{fig:d3}
     \end{subfigure}
    \hfill\\[1em]
    \begin{subfigure}[b]{0.3\textwidth}
         \includegraphics[width=\textwidth]{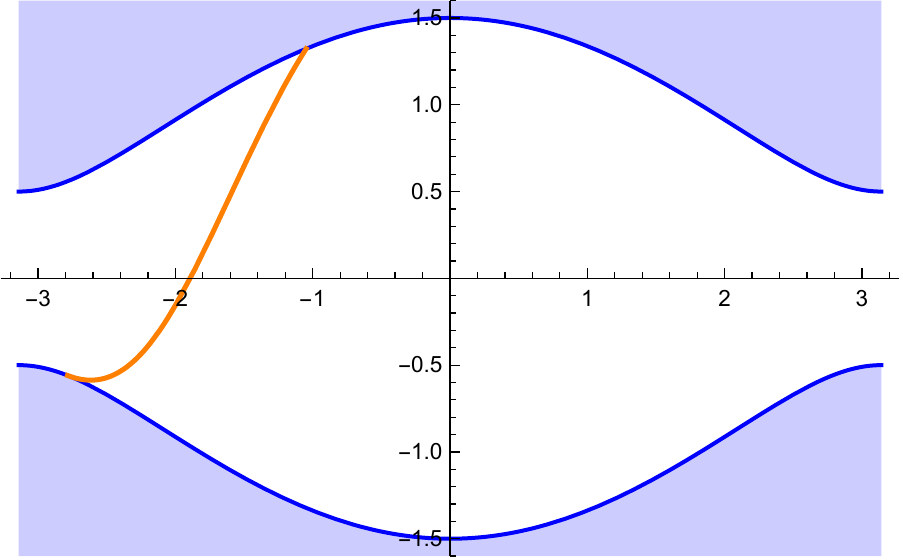}
         \caption{$\theta = 8\pi/7$}
         \label{fig:d4}
     \end{subfigure}
    \hfill    
    \begin{subfigure}[b]{0.3\textwidth}
         \includegraphics[width=\textwidth]{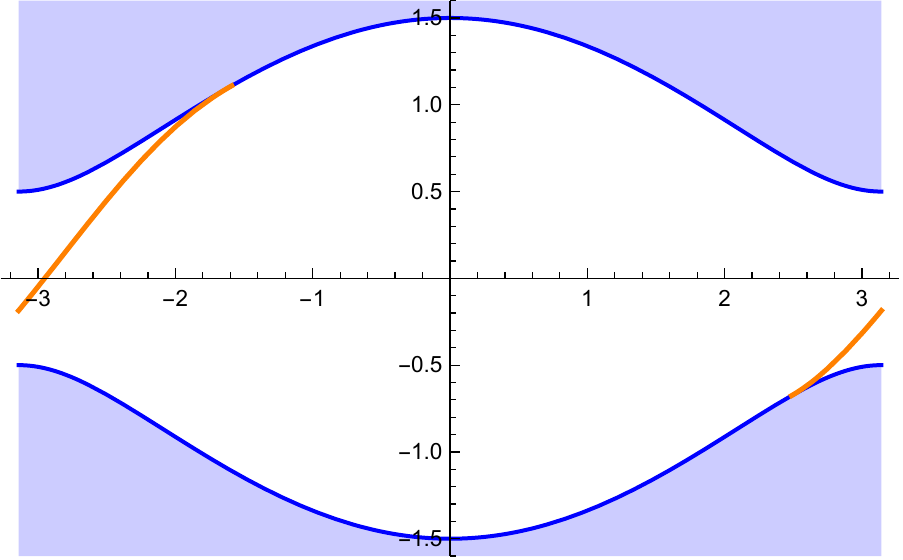}
         \caption{$\theta = 10\pi/7$}
         \label{fig:d5}
     \end{subfigure}
    \hfill
    \begin{subfigure}[b]{0.3\textwidth}
         \includegraphics[width=\textwidth]{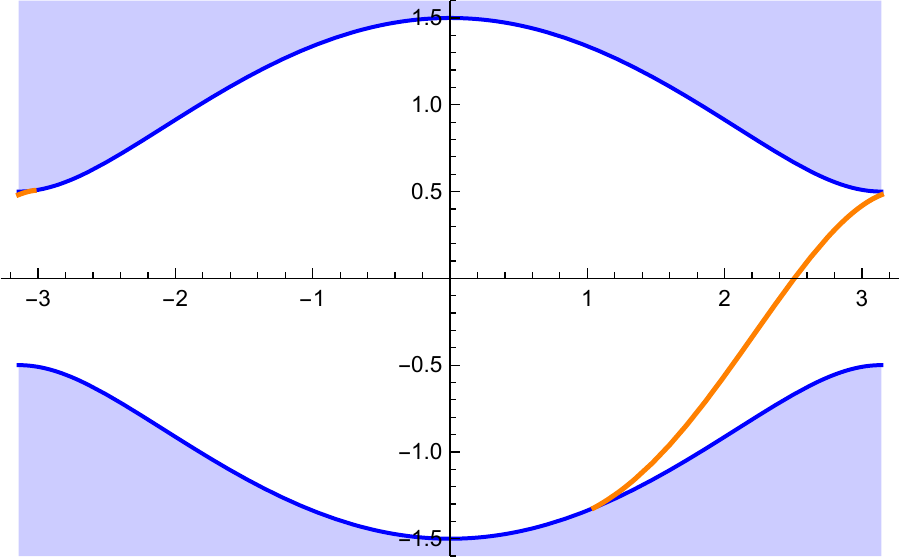}
         \caption{$\theta = 12\pi/7$}
         \label{fig:d6}
     \end{subfigure}
    \hfill    
 \end{center}
 \caption{\raggedright The bulk and edge state dispersion relations of the Wilson fermion model with $k_2=0$ and $m = -1.5$ and the boundary condition parameter $\theta= 2\pi/7$, $4\pi/7$, $6\pi/7$, $8\pi/7$, $10\pi/7$, $12\pi/7$. The horizontal and vertical axes correspond to $k_1$ and $\epsilon$.}
 \label{fig:bulk_edge_disp_2Db}
 \end{figure}
 \begin{figure}[t]
 \begin{center}
    \begin{subfigure}[b]{0.3\textwidth}
         \includegraphics[width=\textwidth]{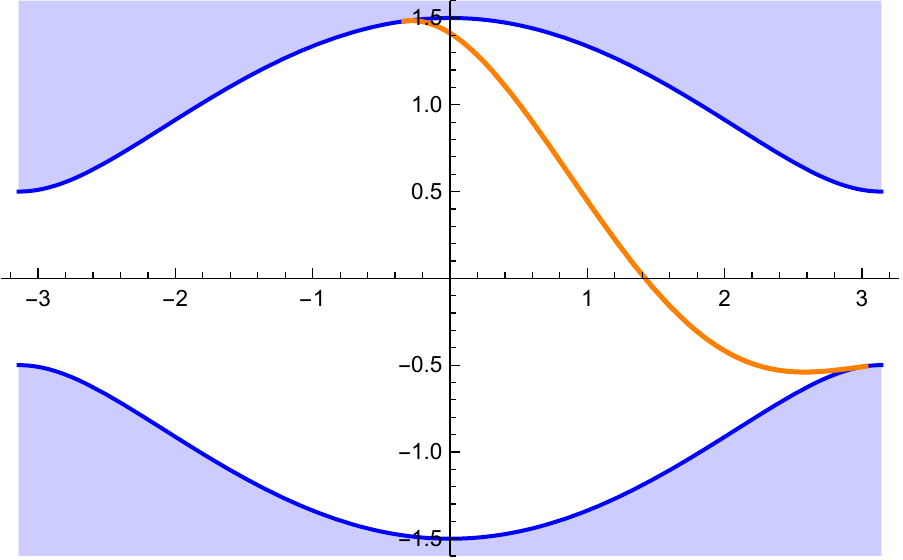}
         \caption{$\theta = 2\pi/7$}
         \label{fig:e1}
     \end{subfigure}
    \hfill
    \begin{subfigure}[b]{0.3\textwidth}
         \includegraphics[width=\textwidth]{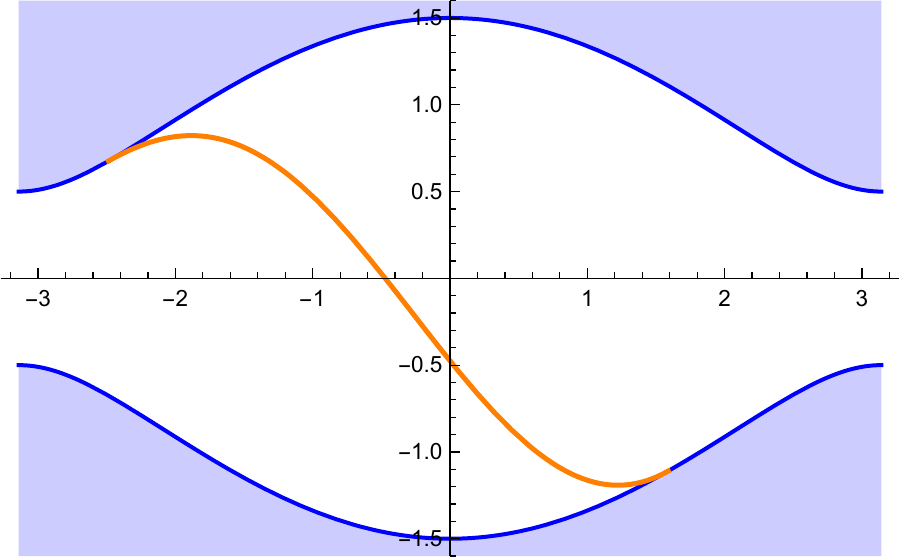}
         \caption{$\theta = 4\pi/7$}
         \label{fig:e2}
     \end{subfigure}
    \hfill    
    \begin{subfigure}[b]{0.3\textwidth}
         \includegraphics[width=\textwidth]{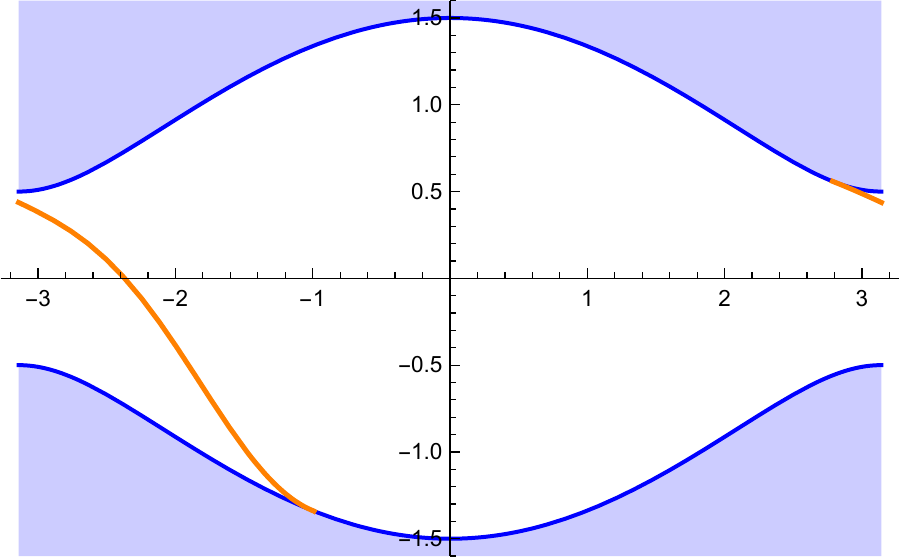}
         \caption{$\theta = 6\pi/7$}
         \label{fig:e3}
     \end{subfigure}
    \hfill\\[1em]
    \begin{subfigure}[b]{0.3\textwidth}
         \includegraphics[width=\textwidth]{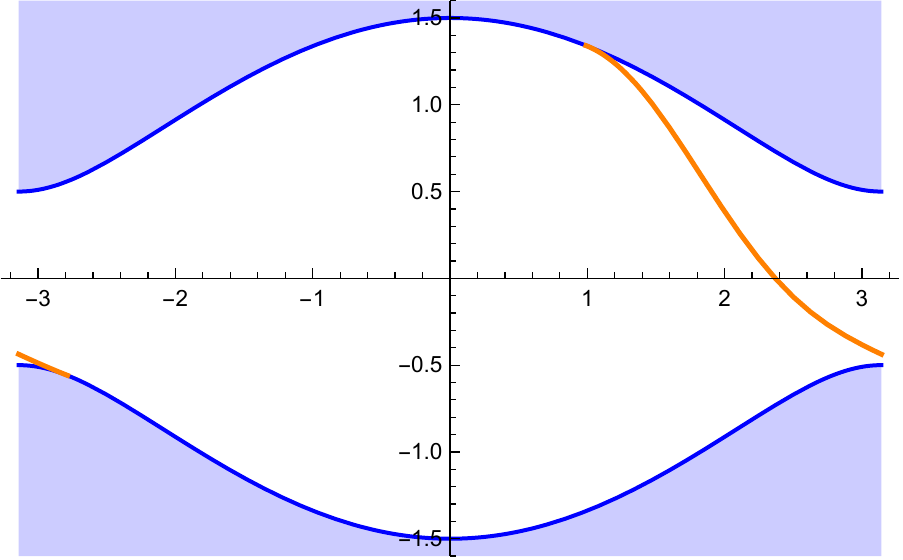}
         \caption{$\theta = 8\pi/7$}
         \label{fig:e4}
     \end{subfigure}
    \hfill    
    \begin{subfigure}[b]{0.3\textwidth}
         \includegraphics[width=\textwidth]{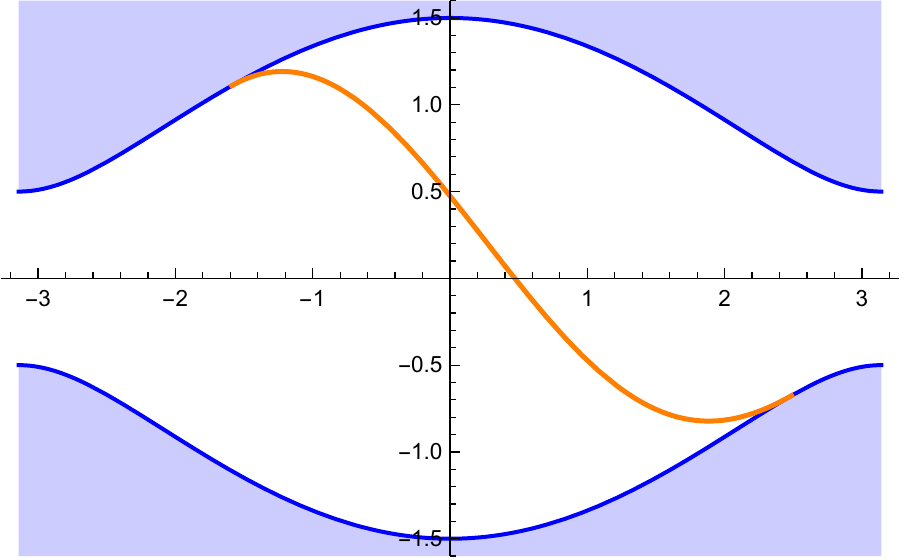}
         \caption{$\theta = 10\pi/7$}
         \label{fig:e5}
     \end{subfigure}
    \hfill
    \begin{subfigure}[b]{0.3\textwidth}
         \includegraphics[width=\textwidth]{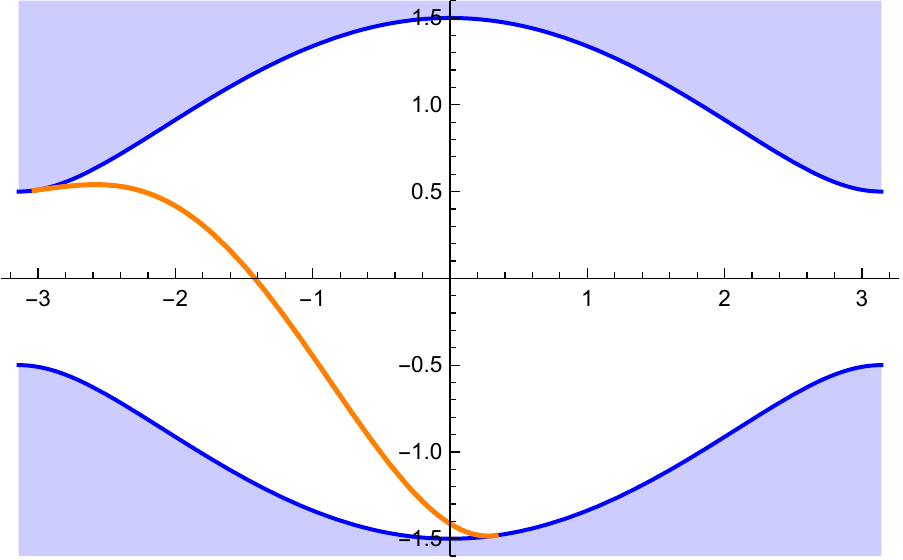}
         \caption{$\theta = 12\pi/7$}
         \label{fig:e6}
     \end{subfigure}
    \hfill   
 \end{center}
 \caption{\raggedright The bulk and edge state dispersion relations of the Wilson fermion model with $k_2=\pi$ and $m = -3.5$ and the boundary condition parameter $\theta= 2\pi/7$, $4\pi/7$, $6\pi/7$, $8\pi/7$, $10\pi/7$, $12\pi/7$. The horizontal and vertical axes correspond to $k_1$ and $\epsilon$.}
 \label{fig:bulk_edge_disp_2Dc}
 \end{figure}
Using this solution $\beta_\pm$, we then obtain the energy eigenvalue, which depends on $(k_1,\theta)$,
\be
	\epsilon_{\pm}=\frac{\cos k_1-m-2\pm\sqrt{A^2+\sin^2 2\theta}}{\sin 2\theta}\,.
\ee
Although we now have apparently two edge states, one of the is not compatible with the normalizability condition $|\beta|<1$.
Hence, we have a single edge state similarly to the continuum theory.
This is shown as follows:  There are two possible edge states in general: $(\epsilon_+, \beta_+)$ if $|\beta_+|<1$ and $(\epsilon_-, \beta_-)$ if $|\beta_-|<1$. 
Considering a function of the parameter $\beta$,
\be
	f(\beta)=\beta^2 \cos^2 \theta +A\beta -\sin^2 \theta\,, \quad f(0)=-\sin^2\theta\le0\,,
\ee
then, the existence of a root $|\beta|<1$ yields
\begin{subequations}\label{be+-}
 \begin{gather}
 \label{be+}
	|\beta_+|<1 \iff f(+1)>0 \iff \cos(k_1+2\theta)>(m+1)\cos 2\theta
	\, , \\
	|\beta_-|<1 \iff f(-1)>0 \iff \cos(k_1+2\theta)<(m+3)\cos 2\theta
	\label{be-}
	\, .
 \end{gather}
\end{subequations}
We remark that these two equations (\ref{be+}) and (\ref{be-}) may be satisfied simultaneously. 
Eqs. \eqref{be+-} in fact provide constraint on the domain of the momentum parameter $k_1$. 
As shown in Figs. \ref{fig:bulk_edge_disp_2Db} and \ref{fig:bulk_edge_disp_2Dc}, we see that the total number of chiral gapless edge state is always one for the various boundary condition parameter $\theta$. 
Hence, we conclude that the chiral edge state is topologically protected in the class A Wilson fermion model even for the generic boundary condition parameter, which does not violates any specific symmetry.

\section{Second order topological insulators and hinge states under lattice boundary conditions}
\label{2OTI}
In this section we consider the boundary condition in two directions, which may describe the second order topological insulator.
We first reformulate our previous results in the continuum model~\cite{Hashimoto:2017ta} with the nontrivial boundary condition in two directions on a lattice, including edge state dispersions, hinge state dispersion.
We then study how a second order topological insulator arises by tuning the boundary conditions.
We demonstrate that the gapless hinge state is protected by the nontrivial topological structure of the gapped edge states.

We start with the following chiral topological insulator model, 
\begin{align}
    \mathcal{H}_S = \sum_k \psi^\dag(k) H_S(k) \psi(k)
    \, , \qquad
	{H}_{S}(k) = \Gamma_5 (2+\sum_{i=1}^3 \cos k_i) + \sum_{i=1}^3 \Gamma_i \sin k_i \,,
\label{HamilS}
\end{align} 
which is a slight modification of that considered in~Ref.~\cite{Schindler2018}.
We remark that this model is obtained from the five-dimensional Weyl semimetal model~\cite{Hashimoto:2020tt}.
We use the following representation of the gamma matrices, following the convention of~\cite{Hashimoto:2017ta}, 
\begin{align}
	\Gamma^i = \left(
	\begin{array}{cc}
	0& -i\sigma_i \\ i\sigma_i & 0
	\end{array}\right) \, , \qquad
	&\Gamma^4 = \left(
	\begin{array}{cc}
	0& \id_2 \\ \id_2 & 0
	\end{array}
	\right) \, , \qquad
	\Gamma^5 = \left(
	\begin{array}{cc}
	\id_2 & 0 \\ 0 & -\id_2
	\end{array}
	\right) \, .
	\label{eq:Gamma_matrices}
\end{align}
We will also use the following Bloch Hamiltonian
\begin{align}
	{ H}_C(k) =R^{-1}{ H}_S(k) R &= \sum_{M=1}^5 \Gamma^M h_M \\
	& = \Gamma_5 \sin k_x + \Gamma_4 \sin k_y - \Gamma_1 \qty(2+\sum_{i=1}^3 \cos k_i) + \Gamma_3 \sin k_z 
	\,,
	\label{HGah}
\end{align} 
which is obtained by the basis rotation through the matrix,
\be
	R=\frac{1}{2}(\id_4-\Gamma_5\Gamma_1)(\id_4-\Gamma_4\Gamma_2)\,.
	\label{eq:R-mat}
\ee
These Hamiltonians exhibit the following chiral symmetries:
\begin{align}
    \{ H_S,\Gamma_S \} = 0 \, , \qquad 
    \{ H_S,\Gamma_C \} = 0 \, , \qquad 
\end{align}
where the corresponding chiral matrices are given by
\begin{align}
    \Gamma_S = \Gamma_4 \, , \qquad
    \Gamma_C = \Gamma_2 \, .
\end{align}



\subsection{Boundary conditions for edge states and hinge states}

In this part we discuss the boundary conditions on a lattice in two directions and their compatibility at the intersection.
We show that the $\cos \hat{k}_i$ terms do not contribute to boundary conditions for edge states and hinge states as in Sec.~\ref{BCL}. 
For the Hamiltonian ${H}_C$, we obtain the following boundary conditions,
\be
	\psi ^{\dagger} \Gamma_5 \psi \Big|_{n_1=1}=0 \,,\quad 
	\psi ^{\dagger} \Gamma_4 \psi \Big|_{n_2=1}=0 \, .
\ee
As discussed in \cite{Hashimoto:2017ta}, these two boundary conditions are solved as follows,
\begin{subequations}\label{eq:bdyn1n2a}
 \begin{align}
	\psi \Big|_{n_1=1} & \propto \left(
	\begin{array}{c}
	\id_2 \\ U_1
	\end{array}
	\right)\xi
	\label{bou55}
	\, , \\
	\psi \Big|_{n_2=1} & \propto 
	\left(
	\begin{array}{c}
	\id_2 -U_2 \\
	\id_2 +U_2
	\end{array}
	\right)\chi \, ,
	\label{x4b} 
\end{align}
\end{subequations}
where $U_1$ and $U_2$ are arbitrary $U(2)$ matrices and $\xi$ and $\chi$ are two arbitrary two-component spinors. 
We may obtain the boundary conditions for $H_S$ by the basis rotation with the matrix $R$ defined in \eqref{eq:R-mat}.
For the later convenience, we may rewrite Eqs.~\eqref{eq:bdyn1n2a} as follows,
\begin{subequations}\label{eq:bdyn1n2b}
 \begin{align}
\left(
\begin{array}{cc}
\id_2 &  -U_1^\dagger 
\\
U_1 & -\id_2
\end{array}
\right) \psi\Big|_{n_1=1} & = 0 \, ,
\label{bou55-+} \\
\left(
\begin{array}{cc}
\frac12(U_2^\dagger-U_2) &  \id_2-\frac12(U_2^\dagger + U_2) 
\\
\id_2 +\frac12( U_2^\dagger + U_2) &  -\frac12(U_2^\dagger-U_2) 
\end{array}
\right) \psi\Big|_{n_2=1} & = 0 \, . \label{bou22}
\end{align}
\end{subequations}
We parametrize the two $U(2)$ matrices as following
\begin{subequations}
 \begin{empheq}[left=\empheqlbrace]{align}
		&U_1 = e^{i\theta_1} U_1'= e^{i\theta_1} (a_0+i\vec{a} \cdot \vec{\sigma})\label{U1the1}\, ,
		\\  
		&U_2 = e^{i\theta_2} U_2'= e^{i\theta_2} (b_0+i\vec{b} \cdot \vec{\sigma})\, .
\end{empheq}
\end{subequations}
We remark that the coefficients obey the constraint $|a|^2 = |b|^2 = 1$, so that we have the decomposition, $e^{i \theta_i} \in U(1)$ and $U_i' \in SU(2)$ for $i = 1,2$~\cite{Hashimoto:2017ta}.
With this parametrization, Eqs.~\eqref{eq:bdyn1n2b} can be further rewritten as 
\begin{subequations}\label{eGamm}
\begin{align}
 \label{eGamm1}
	(e^{i\Gamma_5\theta_1}-a_0\Gamma_4-\vec{a} \cdot \vec{\Gamma})\psi\Big|_{n_1=1} & = 0\,, \\
	(e^{i\Gamma_4\theta_2}+b_0\Gamma_5-\vec{b} \cdot \vec{\Gamma})\psi\Big|_{n_2=1} & = 0\,. \label{eGamm2}
\end{align}
\end{subequations}
These equations are convenient to read off the symmetry.
Both of them are apparently not compatible with the chiral symmetry in general.

In order to consider gapless edge states protected by chiral symmetry generated by $\Gamma_C = \Gamma_2$, we need to consider
\be\label{sin1}
	\theta_i= \qty( n + \frac{1}{2} ) \pi \,, \quad n\in \mathbb{Z}
	\, ,
\ee
which is equivalent to $\cos \theta_i = 0$, and also
\be
	a_2=b_2=0\,.
\ee
If we have non-zero coefficients $(a_2,b_2)$, the chiral symmetry is violated and the edge state is gapped out~\footnote{%
Although the chiral symmetry is also violated in the case $a_2 = b_2 = 0$ with $\cos \theta_i \neq 0$, it is not clear for us at this moment how to construct the hinge state as discussed below.
We leave this issue for a future study.
}.

For any states to exist on the intersection of two boundaries, these two boundary conditions have to be compatible with each other: The wave functions have to satisfy both boundary conditions. 
We consider the compatibility condition of the boundary conditions Eq.~(\ref{x4b}) and Eq.~(\ref{bou55-+}),
\begin{align}
	\left[U_1(\id_2 -U_2)-(\id_2 +U_2)\right]\chi = 0 \, .
\label{consUchi}
\end{align}
This compatibility condition provides constraints for the boundary condition parameters $U_1$ and $U_2$,
\begin{subequations}
 \begin{empheq}[left=\empheqlbrace]{align}
		&\vec{a}\cdot\vec{b}=-\cos \theta_2 \cos \theta_1\, ,
		\\  
		&a_0\sin \theta_2=b_0\sin \theta_1\, .
\end{empheq}
\end{subequations}
Note that this compatibility condition is basis independent. 
\subsubsection{Edge states}
\label{ESD}
We show the dispersion relations of the edge states. 
We consider the edge state wave function in the form of
\be
	\psi_{n_i}=\beta_i^{n_i-1} \psi \Big|_{n_i=1} \, , \qquad  i=1,2 \,,
\ee
with the parameter $\beta_i \in \mathbb{R}$ and $|\beta_i| < 1$ as before.
In the $n_1$ direction, the translation operator exp$(i\hat{k}_1)$ has the eigenvalue $\beta_1$, so that we obtain
\begin{subequations}\label{eq:cosgam1&sinalp1}
\be
	&&\cos \hat{k}_1 \ \longrightarrow \ \frac{1}{2}\qty( \beta_1+\frac{1}{\beta_1})=:\gamma_1
	\label{cosgam1}\\
	&&\sin \hat{k}_1 \ \longrightarrow \ \frac{1}{2i}\qty( \beta_1-\frac{1}{\beta}_1)=:i\alpha_1\label{sinalp1}\,.
\ee
\end{subequations}
In the $n_2$ direction, we instead obtain
\begin{subequations}
 \be
	&&\cos \hat{k}_2 \ \longrightarrow \ \frac{1}{2}\qty( \beta_2+\frac{1}{\beta}_2)=:\gamma_2
	\\
	&&\sin \hat{k}_2 \ \longrightarrow \ \frac{1}{2i}\qty( \beta_2-\frac{1}{\beta}_2)=:i\alpha_2\,.
\ee
\end{subequations}
The remaining part of the calculation is parallel with the continuous model.
We follow the calculation shown in~\cite[Sec.~IIIC]{Hashimoto:2017ta}.
We redefine the coefficients appearing in the Hamiltonian~\eqref{HGah} as follows,
\begin{subequations}
\begin{align}
\left(-i\vec{h} \cdot \vec{\sigma} +h_4\right)U_1' 
 & =: -i\vec{h}^{(1)} \cdot \vec{\sigma} +{h}^{(1)}_4\,,
\label{shu1} \\
\left(+i\vec{h} \cdot \vec{\sigma} +h_5\right)U_2'
 & =: +i\vec{h}^{(2)} \cdot \vec{\sigma} +{h}^{(2)}_5\,,
\label{shu2}
\end{align}
\end{subequations}
from which we obtain\\[1em]
\begin{subequations}
\begin{minipage}{.5\textwidth}
\begin{empheq}[left=\empheqlbrace]{align}
		&{h}^{(1)}_i=a_0h_i-a_ih_4+\epsilon_{ijk}a_jh_k \label{tilhi}\, ,
		\\  
		&{h}^{(1)}_4=a_0h_4+\vec{a}\cdot\vec{h} \label{tilh4}\, ,
\end{empheq}
\end{minipage}
\begin{minipage}{.5\textwidth}
\begin{empheq}[left=\empheqlbrace]{align}
		&{{h}}^{(2)}_i =b_0h_i-b_ih_4+\epsilon_{ijk}b_jh_k\, ,
		\\  
		&{{h}}^{(2)}_5=b_0h_4-\vec{b}\cdot\vec{h} \, .
\end{empheq}
\end{minipage}\\[1em]
\end{subequations}
Using these parametrization, we obtain the energy spectra of the edge states $\epsilon_i$ and the penetration parameters $\alpha_i$ for $i = 1,2$,
\begin{subequations}
\begin{empheq}[left=\empheqlbrace]{align}
		& \epsilon_1=-{h}^{(1)}_4 \cos \theta_1 \pm\sqrt{|\vec{h}^{(1)}|^{2}}\sin \theta_1  \label{e1}\, ,
		\\  
		& \alpha_1=-{h}^{(1)}_4 \sin \theta_1 \mp\sqrt{|\vec{h}^{(1)}|^{2}}\cos \theta_1  \label{alp1} \, .
\end{empheq}
\begin{empheq}[left=\empheqlbrace]{align}
		&   \epsilon_2=-{{h}}^{(2)}_5 \cos \theta_2 \pm\sqrt{|{\vec{h}}^{(2)}|^{2}}\sin \theta_2  \label{e2}\, ,
		\\  
		&  \alpha_2=-{{h}}^{(2)}_5 \sin \theta_2 \mp\sqrt{|{\vec{h}}^{(2)}|^{2}}\cos \theta_2  \label{alp2} \, .
\end{empheq}
\end{subequations}
In Eqs. (\ref{e1}) and (\ref{a1}), the spectrum and the penetration parameter $(\epsilon_1,\alpha_1)$ still depend on $\gamma_1$, namely on $\beta_1$. 
Therefore these two equations are coupled with each other. 
This is the case for Eqs. (\ref{e2}) and (\ref{a2}) as well. 
We can also discuss the dispersion relation as in Sec.~\ref{1OTI} although it could be more complicated.
\subsubsection{Hinge states}
As discussed in \cite{Hashimoto:2017ta}, we have a consistency condition for the dispersion relation of the hinge state $\epsilon(k)$,
\begin{empheq}[left=\empheqlbrace]{align}
	& A\epsilon^2-2B\epsilon+C=0\, ,
 \label{eq:eoe_disp}
	\\ &a_0=b_0=0 \, ,
\end{empheq}
where the coefficients are defined as 
\begin{subequations}
 \label{eq:ABC}
 \begin{align}
  &A := 1-\cos^2\theta_2\cos^2\theta_1  \, , 
  \\
  &B := \vec{a}\cdot\vec{h}\cos\theta_1\sin^2\theta_2+\vec{b}\cdot\vec{h}\cos\theta_2\sin^2\theta_1  \, , 
  \\
  &C := (\vec{a}\cdot\vec{h})^2\sin^2\theta_2+(\vec{b}\cdot\vec{h})^2\sin^2\theta_1- |\vec{h}|^2\sin^2\theta_1\sin^2\theta_2\, .
 \end{align}
\end{subequations}
Imposing the condition~\eqref{sin1}, we have a solution,
\begin{subequations}\label{eq:eoe_coef}
\be
	\epsilon&=&\sqrt{|\vec{h}|^2-(\vec{a}\cdot\vec{h})^2-(\vec{b}\cdot\vec{h})^2}=\vec{c}\cdot\vec{h}\label{eoedis}\\
	\alpha_1&=&-\vec{a}\cdot\vec{h} \label{eoea1}\\
	\alpha_2&=&-\vec{b}\cdot\vec{h} \label{eoea2}
\ee
\end{subequations}
for some $\vec{c}$ such that $\vec{a}\cdot\vec{c}=\vec{b}\cdot\vec{c}=0$. 
We will show in the next part that gapless hinge states are realized by a further tuning the coefficients $a_i$ and $b_i$. 

%
 
\subsection{Construction of second order topological insulator from boundary conditions}

In order to realize the second-order topological insulator, we require the following conditions: (i) the edge state is gapped, (ii) the edge state has a nontrivial topological number, and (iii) the hinge state is gapless.
In this part, we discuss how to impose these conditions using the boundary conditions.
Based on the discussion above, we consider $\cos \theta_i = 0$ and $a_0 = b_0 = 0$ in this part. 
Then, the boundary condition (\ref{eGamm}) becomes
\begin{subequations}\label{eGamm_sim}
\be\label{eGamm1sim}
	(i\Gamma_5-\vec{a}\cdot\vec{\Gamma})\psi\Big|_{n_1=1} = 0\,, \\
    \label{eGamm2sim}
	(i\Gamma_4-\vec{b}\cdot\vec{\Gamma})\psi\Big|_{n_2=1} = 0\,.
\ee
\end{subequations}

\subsubsection{Gapped edge states}
In this case, the energy spectra of the edge states (\ref{e1}) and (\ref{e2}) are given by
\be\label{e1e2}
	\epsilon_1=\sqrt{h_4^2+|\vec{a}\times\vec{h}|^2} \,, \quad \epsilon_2=\sqrt{h_5^2+|\vec{b}\times\vec{h}|^2} \,.
\ee
We first consider the continuum limit of the Hamiltonian~\eqref{HGah} for simplicity,
\be
	H_C \ \longrightarrow \ {H}_{CC}= k_x\Gamma_5+ k_y\Gamma_4-m\Gamma_1+k_z\Gamma_3
	\, .
\ee
Then, the energy spectra (\ref{e1e2}) are given by
\begin{subequations}
\be\label{e1e2cc}
	\epsilon_1=\sqrt{k_y^2+(a_2m)^2+(a_2k_z)^2+(a_3m+a_1k_z)^2}\,,\\ 
	\epsilon_2=\sqrt{k_x^2+(b_3m)^2+(b_3k_z)^2+(b_3m+b_1k_z)^2}\,.
\ee
\end{subequations}
From these expressions, we see that the edge spectra are gapless if and only if $a_2=0$, $b_2=0$ respectively, in which the chiral symmetry is preserved on the boundary. 
Considering the original lattice model, the energy spectra (\ref{e1e2}) are given by
\begin{subequations}
\be\label{e1e2l}
	\epsilon_1=\sqrt{\sin k_y^2+(a_2 h_1)^2+(a_2 \sin k_z)^2+(a_3h_1-a_1\sin k_z)^2}\,,
	\\ \epsilon_2=\sqrt{\sin k_x^2+(b_2h_1)^2+(b_2\sin k_z)^2+(b_3h_1-b_1\sin k_z)^2}\,.
\ee
\end{subequations}
Here, the coefficient $h_1$ should play a role of the mass parameter in the lattice model, which depends on $\gamma_i$ for each edge state.
In fact, we can show that the coefficient $h_1$ is non-vanishing at the possible gapless points, $\sin k_y=\sin k_z=0$ for $\epsilon_1$, as follows,
\be
	\alpha_1\Big|_{\sin k_y=\sin k_z=0}=-{h}^{(1)}_4\Big|_{\sin k_y=\sin k_z=0}=-a_1h_1\Big|_{\sin k_y=\sin k_z=0}
\ee
where for the edge states we have
\be
	\alpha_1=\frac{1}{2}(\beta^{-1}_1-\beta_1)\ne0 \, .
\ee
The argument is the same for the other case $\epsilon_2$. 
Therefore, in the case of lattice model, we have the same conclusion as in the continuum limit: In order to gap out the edge states, we need to violate the chiral symmetry on the boundary.
\subsubsection{Boundary conditions and topological number of edge states}
We calculate a topological number of the edge states in this part. 
The normalized edge state wave function depending on the boundary condition (\ref{bou55}) is given by
\be\label{edgewave}
	\psi_{n_1}=\frac{1}{\sqrt{2}}\left(\begin{array}{c}1 \\U_1\end{array}\right)\xi \sqrt{1-\beta^2}\beta^{n_1}\,,
\ee
where we also normalize the spinor $\xi$ satisfing Eq.~(\ref{equi2}), as
\be
	\xi^{\dagger}\xi=1\,.
\ee
We define the Berry connection for the edge state in a similar way as in the continuum theory~\cite{Hashimoto:2016dtm}:
\be\label{Bercon}
	\vec{A}&=&i\sum_{n_1 \ge 1}\psi_{n_1}^{\dagger}\frac{\partial}{\partial \vec{k}} \psi_{n_1}\,.
\ee
From the wave function~(\ref{edgewave}), we obtain 
\be
	\vec{A}&=&i\sum_{n_1 \ge 1}  \left(\frac{1}{\sqrt{2}}\left(\begin{array}{c}1 \\U_1\end{array}\right)\xi \sqrt{1-\beta^2}\beta^{n_1}\right)^{\dagger}
	\frac{\partial}{\partial \vec{k}} \left(\frac{1}{\sqrt{2}}\left(\begin{array}{c}1 \\U_1\end{array}\right)\xi \sqrt{1-\beta^2}\beta^{n_1}\right)\nonumber\\
	&=&i \xi^{\dagger}  \frac{\partial}{\partial \vec{k}}\xi +i\sum_{n_1 \ge 1} \sqrt{1-\beta^2}\beta^{n_1}\frac{\partial}{\partial \vec{k}} ( \sqrt{1-\beta^2}\beta^{n_1} )\nonumber\\
	&=& i \xi^{\dagger}  \frac{\partial}{\partial \vec{k}}\xi\,.
\ee
Considering the boundary condition parameter $ a_2\ne 0 $, we obtain a gapped spectrum from Eq.~(\ref{equi2}).
In this case, we will have a Chern number which can be written in terms of coefficients of the effective Hamiltonian in Eq. (\ref{equi2}),
\be
    H_{\text{eff}}=-\alpha_1\cot\theta_1-\frac{\vec{h}^{(1)}\cdot\vec{\sigma}}{\sin\theta_1}\,.
\ee
Then, the topological number is calculated as
\be
	N_1 
	&=&\int \frac{d k}{2\pi} (\partial_1 A_2-\partial_2 A_1)\nonumber\\
	&=&\frac{1}{8\pi } \int \epsilon_{ijk} \frac{{h}^{(1)}_i}{|{h}^{(1)}|^3} d {h}^{(1)}_j \wedge d {h}^{(1)}_k\,.\label{Chhhh}
\ee
Putting $h_2 = 0$, the coefficients (\ref{tilhi}) become
\begin{subequations}
\be
	{h}^{(1)}_1&=&-a_1h_4+a_2h_3 \, , \\
	{h}^{(1)}_2&=&-a_2h_4+a_3h_1-a_1h_3 \, , \\	
	{h}^{(1)}_3&=&-a_3h_4-a_2h_1 \, .
\ee
\end{subequations}
Hence, we can obtain a nonzero Chern number if $a_2 \neq 0$.
so we can see that $a_2$ indeed should be nonzero to give a nontrivial Chern number. 
This argument is also applied for the Chern number $N_2$ associated with the edge state localized on the boundary $n_2=1$. 

We demonstrate to obtain a nonzero Chern number. 
For this purpose, we may apply the formula~\cite{Sticlet:2012wl}: 
\be\label{N1sgnh}
	N_1=\frac{1}{2} \sum_{k^{(a)}}\text{sgn}({h}^{(1)}_3)
	\left.
	\operatorname{sgn}
	\begin{pmatrix}
	\partial_z h^{(1)}_1 & \partial_z h^{(1)}_2 \\
	\partial_y h^{(1)}_1 & \partial_y h^{(1)}_2
	\end{pmatrix}
    \right|_{k_i=k_i^{(a)}}
\ee
where $k_i^{(a)}$ are points in Brillouin zone at which ${h}^{(1)}_1(k^{(a)})={h}^{(1)}_2(k^{(a)})=0$, and the derivatives are defined as $\partial_y = \partial/\partial k_y$, $\partial_z = \partial/\partial k_z$.
For a matrix $A$, we define $\operatorname{sgn}(A) = \operatorname{sgn}(\det A)$. 
 Taking $a_3=0$ for simplicity, Eq. (\ref{N1sgnh}) becomes
 \be
	N_1&=&\frac{1}{2} \sum_{k^{(a)} :~\sin k_y^{(a)}=\sin k_z^{(a)}=0}\text{sgn}(a_2h_1)\text{sgn}(\cos k_y\cos k_z)\nonumber\\
	&=&\frac{1}{2} \text{sgn}(a_2)\qty(\text{sgn}(h_1)\Big|_{(k_y,k_z)=(0,0)}+\text{sgn}(h_1)\Big|_{(k_y,k_z)=(\pi,\pi)}-\text{sgn}(h_1)\Big|_{(k_y,k_z)=(0,\pi)}-\text{sgn}(h_1)\Big|_{(k_y,k_z)=(\pi,0)})\nonumber\\
	&=&\text{sgn}(a_2)\,.
\ee
The calculation of $N_2$ is completely parallel to $N_1$ and we  get, for $b_3=0$
 \be
	N_2&=&\frac{1}{2} \sum_{k^{(a)}}\text{sgn}({{h}}^{(2)}_3)
	\left.
	\operatorname{sgn}
	\begin{pmatrix}
	\partial_x h^{(2)}_1 & \partial_x h^{(2)}_2 \\
	\partial_z h^{(2)}_1 & \partial_z h^{(2)}_2
	\end{pmatrix}
    \right|_{k_i=k_i^{(a)}}\nonumber
	\\&=&\frac{1}{2} \sum_{k^{(a)}:~ \sin k_x^{(a)}=\sin k_z^{(a)}=0}\text{sgn}(b_2h_1)\text{sgn}(\cos k_x\cos k_z)\nonumber\\
	&=&\frac{1}{2} \text{sgn}(b_2)\qty(\text{sgn}(h_1)\Big|_{(k_x,k_z)=(0,0)}+\text{sgn}(h_1)\Big|_{(k_x,k_z)=(\pi,\pi)}-\text{sgn}(h_1)\Big|_{(k_x,k_z)=(0,\pi)}-\text{sgn}(h_1)\Big|_{(k_x,k_z)=(\pi,0)})\nonumber\\
	&=&\text{sgn}(b_2)\,.
\ee
See Appendix \ref{Comsgnh} for details of the computation.

\subsubsection{Gapless hinge states}
We find out the hinge state dispersion relation and the corresponding wave function in the case $a_3=b_3=0$, for which the topological numbers are obtained.
In this case, we have the relations for the coefficients, $b_1=\mp a_2, b_2=\pm a_1$ and $c_1=c_2=0,~c_3=\pm1$.
Then, from Eqs.~\eqref{eq:eoe_coef}, we obtain the gapless spectrum,
\begin{subequations}
\be
	\epsilon&=&\pm\sin k_z \label{eoedis1}\\
	\alpha_1&=&-a_1h_1 \label{eoea11}\\
	\alpha_2&=&-b_1h_1\label{eoea21}\,.
\ee
\end{subequations}
We check the normalizability of the wave function.   
The parameter relations $\alpha_i=\frac{1}{2}(\frac{1}{\beta_i}-\beta_i)$ for $i = 1,2$ can be rewritten as
\be
	\beta^2_i+2\alpha_i\beta_i-1=0\,,
\ee
where no summation over the index $i$. 
Since the discriminant of this quadratic equation is given by $\Delta =4\alpha_i^2+4=:4\gamma_i^2$, the reality of $\gamma_i$ guarantees the reality of $\beta_i$.
Furthermore, because of the normalizability condition $|\beta_i|<1$, we have a one-to-one correspondence: a positive $\alpha_i$ corresponds to a positive $\beta_i$ and a negative $\alpha_i$ corresponds to negative $\beta_i$. 
Hence, it is sufficient to determine the parameter $\alpha_i$. 
We have consistent solutions for 
\begin{subequations}\label{eoea_k}
\be
	\alpha_1&=&-a_1(2+\gamma_1+\gamma_2+\cos k_z) \label{eoea1k} \, , \\
	\alpha_2&=&-b_1(2+\gamma_1+\gamma_2+\cos k_z)\label{eoea2k}\,.
\ee
\end{subequations}
Noticing that $\alpha^2_i=\gamma^2_i-1$ and defining 
\begin{subequations}
\begin{empheq}[left=\empheqlbrace]{align}
	& \gamma_{+}=\frac{\gamma_1+\gamma_2}{2}\, ,
	\\ &\gamma_{-}=\frac{\gamma_1-\gamma_2}{2} \, ,
\end{empheq}
\end{subequations}
we obtain the following relations from Eq.~\eqref{eoea_k},
\begin{subequations}\label{ga+ga-1&2}
\be
	(\gamma_++\gamma_-)^2-1&=&a_1^2(2+\cos k_z+2\gamma_+)^2 \label{ga+ga-1} \, , \\
	(\gamma_+-\gamma_-)^2-1&=&b_1^2(2+\cos k_z+2\gamma_+)^2\label{ga+ga-2}\,.
\ee
\end{subequations}
Obtaining the relation from \eqref{ga+ga-1&2} 
\begin{subequations}
\be
	2\gamma^2_++2\gamma^2_--2&=&(2+\cos k_z+2\gamma_+)^2 \, , \\
	4\gamma_+\gamma_-&=&(a_1^2-b_1^2)(2+\cos k_z+2\gamma_+)^2 \,,
\ee
\end{subequations}
we eliminate the variable $\gamma_-$ to obtain a quartic equation for the variable $\gamma_+$,
\be\label{fga+}
	f(\gamma_+) = 0 \, .
\ee 
The function $f(z)$ is defined as
\begin{align}
    f(z) := 8z^2((2 z + 2 + \cos k_z)^2-2 z^2 + 2)-(a_1^2-b_1^2)^2(2 z +2+\cos k_z)^2=0\,.
\end{align}
showing the following behaviors,
\be
	\lim_{z\to \pm \infty }f(z) \to +\infty \,, \qquad  
	f(0)=-(a_1^2-b_1^2)^2(2+\cos k_z)^2\le 0\,.
\ee
Therefore, there are at least two real solutions to Eq. (\ref{fga+}), which are possibly degenerated at $a_1=b_1$. These two solutions of $\gamma_+$ gives two pairs of $\alpha_1, \alpha_2$ in Eqs.~\eqref{eoea_k}, which ensure two normalizable wave functions.

We discuss the relation between the edge state topology and the hinge states. 
If the edge state localized on the boundary $n_1=1$ is topologically trivial $N_1 = 0$, we have $a_2=0$, which shows
\be
	\alpha_2=\mp a_2h_1=0\,.
\ee
This means $\beta_2=\pm1$, so that the wave function is not normalizable in the $n_1$ direction; 
It is not localized on the boundary.
From this point of view, we establish the correspondence between the topologically nontrivial gapped edge state and the normalizable gapless hinge state.

\section{Summary and discussion}
\label{sum}
We have shown that the Hermiticity of the difference operator in the presence of the boundary provides the boundary conditions of topological materials, from which we can further determine the dispersion relation of the edge states. 
We have shown that the properties of the lattice model are consistent with the continuum model for the localised edge/hinge states.

We have analytically studied three lattice models: one-dimensional SSH model in class AIII, two-dimensional Wilson fermion model in class A, and three-dimensional chiral topological insulator model in class AIII. 
In order to have a gapless edge state, the boundary condition should respect the symmetry of the original Hamiltonian if it exists. 
Hence, we need the constraint for the boundary condition in the case of class AIII, while no constraint is necessary for class A, which does not exhibit a specific symmetry.
From this point of view, it would be interesting to generalize the analysis in this paper to other symmetry classes, and study the compatible boundary condition associated with the corresponding symmetry.

We have pointed out that the compatibility of the boundary condition plays a crucial role to have the higher order topological insulator.
We have shown that for the second order topological insulators to exist, the boundary conditions have to violate the symmetries of the bulk Hamiltonian to gap out the edge state. 
Moreover, the boundary conditions are involved in topological charges of the edge states, so that they affect the topological charges of edge states and the structure of hinge states. 
We remark that the second order topological insulator that we construct is classified into the extrinsic high order topological insulators rather than the intrinsic ones~\cite{Okugawa2019}.

\subsection*{Acknowledgments}

The work of TK was supported in part by ``Investissements d'Avenir'' program, Project ISITE-BFC (No.~ANR-15-IDEX-0003), EIPHI Graduate School (No.~ANR-17-EURE-0002), and Bourgogne-Franche-Comté region.


\appendix


\section{Generality of edge state wave function}
\label{sec:gen edge}
We show that in general the edge wave function is written in the form of \eqref{eq:edge_wf}.
Introducing an extended wave function with respect to the sublattice structure
\be
	\Psi_n=\left(\begin{array}{c}\psi_{2n-1} \\\psi_{2n}\end{array}\right) \, ,
\ee
we may rewrite the boundary condition \eqref{bc01} in a local form. 
Then, the edge state wave function will be given by $\Psi_{n+1}=\tilde{\beta}\Psi_n$, and one can show that $\psi_{n+1}=\beta\psi_n$ with $\beta^2=\tilde{\beta}$, which ensures the generality of the edge state wave function \eqref{eq:edge_wf}.

For this purpose, we first rewrite the kinetic terms in terms of the extended wave function,
\begin{subequations}
\begin{align}
	\sum_{m=1}^{2N} \psi_m^{\dagger}\cos \hat{k} \, \psi_m & =\sum_{n=1}^{N} \Psi_n^{\dagger}\left(\begin{array}{cc}0 & 1+\frac{1}{2}\nabla^{\dagger} \\ 1+\frac{1}{2}\nabla & 0\end{array}\right)\Psi_n \, , \\
	\sum_{m=1}^{2N} \psi_m^{\dagger}\sin \hat{k} \, \psi_m & =\sum_{n=1}^{N} \Psi_{n}^{\dagger}\left(\begin{array}{cc}0 & \frac{i}{2}\nabla^{\dagger} \\ -\frac{i}{2}\nabla & 0\end{array}\right)\Psi_n \, ,
\end{align}
\end{subequations}
where we define the difference operator as before,
\begin{subequations}
\be
	\nabla\Psi_n=\Psi_{n+1}-\Psi_n \, ,
	\\
	\nabla^{\dagger}\Psi_n=\Psi_{n-1}-\Psi_n\,.
\ee
\end{subequations}
Then, we consider the Hamiltonian
\be
	\mathcal{H} = \sum_{m=1}^{2N}  \psi_m^{\dagger}((h_1+\cos \hat{k})\sigma_1+\sin \hat{k}\sigma_2+h_3\sigma_3)\psi_m
	\, ,
\ee
which describes the Wilson fermion when the coefficients are given by $h_1=M+\cos \hat{k}'$ and $h_3=\sin \hat{k}'$, and the SSH model when $h_1=s$ and $h_3=0$.
We may rewrite this Hamiltonian as follows,
\be
	\mathcal{H} = \sum_{n=1}^{N}\Psi_{n}^{\dagger}\left((h_1\sigma_1+h_3\sigma_3)\otimes \id_2+\left(\begin{array}{cc}0 & 1+\frac{1}{2}\nabla^{\dagger} \\ 1+\frac{1}{2}\nabla & 0\end{array}\right) \otimes\tau_1 +\left(\begin{array}{cc}0 & \frac{i}{2}\nabla^{\dagger} \\ -\frac{i}{2}\nabla & 0\end{array}\right)\otimes \tau_2 \right) \Psi_{n}
	\,.\nonumber\\
\ee
where $(\tau_i)_{i = 1, 2, 3}$ are the Pauli matrices with respect to the sublattice structure.
Denoting $\hat{P}=2\hat{k}$, we explicitly have
\be
	\left(\begin{array}{cc}0 & 1+\frac{1}{2}\nabla^{\dagger} \\ 1+\frac{1}{2}\nabla & 0\end{array}\right) \otimes\tau_1+\left(\begin{array}{cc}0 & \frac{i}{2}\nabla^{\dagger} \\ -\frac{i}{2}\nabla & 0\end{array}\right)\otimes \tau_2
	=\left(\begin{array}{cccc}
	0 & 0 & 0 & e^{-i\hat{P}} 
	\\0 & 0 & 1 & 0 
	\\0 & 1 & 0 & 0 
	\\e^{i\hat{P}} & 0 & 0 & 0\end{array}\right)\, .
\ee

We consider the eigenvalue equation for this Hamiltonian with the relation $\Psi_2=\tilde{\beta}\Psi_1$:
\begin{subequations}
\begin{empheq}[left=\empheqlbrace]{align}
	&(-\epsilon+h_1\sigma_1+h_3\sigma_3))\psi_1+\left(\begin{array}{cc}0 &  \tilde{\beta}^{-1} \\1 & 0\end{array}\right) \psi_2=0 \, , \label{ep1p2}
	\\ &\left(\begin{array}{cc}0 & 1 \\ \tilde{\beta} & 0\end{array}\right)\psi_1
	+(-\epsilon+h_1\sigma_1+h_3\sigma_3))\psi_2=0\label{p1ep2} \, ,
\end{empheq}
\end{subequations}
which can be rewritten as 
\begin{subequations}
\begin{empheq}[left=\empheqlbrace]{align}
	&\psi_2=-\left(\begin{array}{cc}0 &  1 \\ \tilde{\beta} & 0\end{array}\right) (-\epsilon+h_1\sigma_1+h_3\sigma_3)\psi_1 \label{p2betap1}
	\\ & \left(\id_2-\left(\begin{array}{cc}0 &  \tilde{\beta}^{-1} \\1 & 0\end{array}\right) (-\epsilon+h_1\sigma_1+h_3\sigma_3)\left(\begin{array}{cc}0 &  1 \\ \tilde{\beta} & 0\end{array}\right) (-\epsilon+h_1\sigma_1+h_3\sigma_3)\right)\psi_1=0\label{e2p1} \, .
\end{empheq}
\end{subequations}
We first solve Eq. (\ref{e2p1}). 
Denoting $\tilde{\beta}=\beta^2$, the energy eigenvalue is given by
\be
	\epsilon^2&=&h^2_3+h^2_1\pm\sqrt{h^2_1(2+ \tilde{\beta}^{-1}+ \tilde{\beta})}\nonumber
	\\
	&=&h^2_3+h^2_1\pm h_1(\beta+\beta^{-1}) \, . \label{ee2}
\ee 
Hence, there are four energy eigenvalues as follows,
\begin{subequations}
\be
	e_{1\pm}=\sqrt{h^2_3+h^2_1\pm h_1(\beta+\beta^{-1})}\\
	e_{2\pm}=-\sqrt{h^2_3+h^2_1\pm h_1(\beta+\beta^{-1})}
\ee
\end{subequations}
Substituting Eq. (\ref{ee2}) into Eq. (\ref{e2p1}), we obtain the wave function
\be\label{eigenp1}
	\psi_1\propto  \left(\begin{array}{c}h_1\pm \beta^{-1} \\ \epsilon_{1\pm} - h_3\end{array}\right) \text{ or } \left(\begin{array}{c}h_1\pm \beta^{-1} \\ \epsilon_{2\pm} - h_3\end{array}\right)
\ee
Substituting Eq.(\ref{eigenp1}) back into Eq.(\ref{p2betap1}) with some algebras, we obtain
\be
	\psi_2=\pm\beta \psi_1\,,
\ee
which confirms the generality of the wave function~\eqref{eq:edge_wf}.

\section{Computation of dispersion relations for edge/hinge states }

The edge state solution to the bulk Hamiltonian eigenvalue equation, which is associated with the boundary $n_1 = 1$, is given by
\begin{align}
\psi_{n_1} = \left(
\begin{array}{c}
\xi \\ \eta
\end{array}
\right)\beta_1^{n_1} \,, \quad 
\alpha_1 := \sqrt{-\epsilon^2  + |\vec{h}|^2 +h_4^2} \, .
\end{align}
Together with the boundary conditions, Eq.~(\ref{bou55}) and Eq.~(\ref{HGah}),
the Hamiltonian eigenvalue equation can be written as
\begin{subequations}\label{eq:a1&2}
\begin{align}
\left[(i\alpha_1-\epsilon) + \left(-i \vec{h} \cdot \vec{\sigma} +h_4\right)U'_1\right]\xi & = 0  \, ,
\label{a1}\\
\left[-(i\alpha_1+\epsilon)U'_1 + \left(i\vec{h} \cdot \vec{\sigma} +h_4\right)\right]\xi & = 0  \, .
\label{a2}
\end{align}
\end{subequations}
for $i=1,2,3$.
The boundary condition parameters rotate the coefficients in the Hamiltonian $(h_i)_{i = 1, \ldots,4}$,
\begin{align}
\left(-i\vec{h} \cdot \vec{\sigma} +h_4\right)U_1'
=-i\vec{h}^{(1)} \cdot \vec{\sigma} +{h}^{(1)}_4
\label{tp}
\end{align}
with the condition
\begin{align}
|\vec{h}|^2 + h_4^2 = |\vec{h}^{(1)}|^2 + {h_4^{(1)}}^2 \, .
\end{align}
Then, we may rewrite the two equations \eqref{eq:a1&2} in terms of the rotated coefficients,
\begin{subequations}
\begin{align}
\left[e^{-i\theta_1}(i\alpha_1-\epsilon) +{h}^{(1)}_4-i \vec{h}^{(1)} \cdot {\vec{\sigma}}
\right]\xi & = 0  \, ,\\
\left[-e^{i\theta_1}(i\alpha_1+\epsilon)+{h}^{(1)}_4+i\vec{h}^{(1)} \cdot {\vec{\sigma}}\right]\xi & = 0  \, .
\end{align}
\end{subequations}
which can be equivalently written as
\begin{subequations}
\begin{align}
\left[\alpha_1\sin\theta_1-\epsilon\cos\theta_1 +{h}^{(1)}_4
\right]\xi = 0  \, ,
\label{equi1}\\
\left[\alpha_1\cos\theta_1+\epsilon\sin\theta_1-\vec{h}^{(1)} \cdot {\vec{\sigma}}\right]\xi = 0  \, .
\label{equi2}
\end{align}
\end{subequations}
We have the compatibility condition for these equations,
\begin{subequations}
\begin{align}\label{alp1sinep1cos}
\alpha_1\sin\theta_1-\epsilon\cos\theta_1 +{h}^{(1)}_4 = 0  \, ,\\
\det\left[\alpha_1\cos\theta_1+\epsilon\sin\theta_1-\vec{h}^{(1)} \cdot {\vec{\sigma}}\right]= 0  \, ,
\end{align}
\end{subequations}
where the second equation implies
\begin{align}\label{alp1cosep1sin}
\alpha_1\cos\theta_1+\epsilon\sin\theta_1 = \pm\sqrt{ | \vec{h}^{(1)} |^2} \, .
\end{align}
Therefore, from \eqref{alp1sinep1cos} and \eqref{alp1cosep1sin}, we obtain the dispersion relation of the edge state localized on the boundary $n_1 = 1$,
\begin{subequations}
\begin{align}
\epsilon & = {h}^{(1)}_4\cos\theta_1 \pm \sqrt{|\vec{h}^{(1)}|^2} \sin\theta_1 \, , \\
\alpha_1 & = -{h}^{(1)}_4\sin\theta_1 \pm \sqrt{|\vec{h}^{(1)}|^2} \cos\theta_1 \, .
\label{al5}
\end{align}
\end{subequations}

Next we solve the hinge state eigenvalue equations from Eqs. (\ref{x4b}) and  (\ref{HGah}), 
\begin{subequations}\label{eq:a1&a2--}
\begin{align}
\left[(i\alpha_2-\epsilon) + \left(-i\vec{h} \cdot \vec{\sigma} -i\alpha_1\right)U_2\right]\chi = 0  \, ,
\label{a1--}\\
\left[-(i\alpha_2+\epsilon)U_2+ \left(i\vec{h} \cdot \vec{\sigma} -i\alpha_1\right)\right]\chi = 0  \, .
\label{a2--}
\end{align}
\end{subequations}
where we define
\begin{align}
	h_5:=i\alpha_1 \label{pa5}\,.
\end{align}
Similarly to the previous case (\ref{tp}), we have the new coefficients,
\begin{align}
	(i\vec{h} \cdot \vec{\sigma}+h_5)(b_0+i\vec{b} \cdot \vec{\sigma})=i\vec{h}^{(2)} \cdot \vec{\sigma}+{{h}}^{(2)}_5\, ,
\end{align}
from which we read
\begin{subequations}
\begin{empheq}[left=\empheqlbrace]{align}
	&{{h}}^{(2)}_5=b_0h_5-\vec{b} \cdot \vec{h} \,,
	\\
	&{{h}}^{(2)}_i=b_0h_i+b_ih_5+\epsilon_{ijk}b_jh_k\, .
\end{empheq}
\end{subequations}
Then, we have the solution to \eqref{eq:a1&a2--} as follows,
\begin{subequations}
\begin{empheq}[left=\empheqlbrace]{align}
	&\epsilon \cos\theta_2-\alpha_2\sin \theta_2+{{h}}^{(2)}_5=0\, ,
	\\
	&(\epsilon \sin\theta_2+\alpha_2\cos\theta_2)^2-|\vec{h}^{(2)}|^2=0\, .
\end{empheq}
\end{subequations}
which are equivalent to 
the following set of equations, 
\begin{subequations}
\begin{empheq}[left=\empheqlbrace]{align}
	&\epsilon \cos\theta_2-\alpha_2\sin \theta_2=\vec{b} \cdot \vec{h}-b_0h_5\, ,
	\\
	&\epsilon^2=|\vec{h}|^2-\alpha_2^2-\alpha_1^2 \,.  \label{epaa}
\end{empheq}
\end{subequations}
As $h_5$ is imaginary from the definition \eqref{pa5},
these are three real equations including
 $	b_0h_5=0$, 
 which implies 
  \begin{align}
 	b_0=0\, ,
 \end{align}
 and 
 \begin{align}\label{disp4}
 	\epsilon \cos\theta_2-\alpha_2\sin \theta_2=\vec{b} \cdot \vec{h} . 
 \end{align}

Similarly, we consider the boundary condition in the $n_2$ direction. 
Putting 
\begin{align}
	h_4=i\alpha_2
	\, ,
\end{align}
and from Eq. (\ref{alp1sinep1cos}), we obtain the following relations,
\begin{subequations}
 \begin{empheq}[left=\empheqlbrace]{align}
	&\epsilon \cos\theta_1-\alpha_1\sin \theta_1=\vec{a} \cdot \vec{h}\, , \label{disp5}
	\\ &a_0=0 \, .
\end{empheq}
\end{subequations}
Combining the relations \eqref{disp4} \eqref{disp5} and  \eqref{epaa} to eliminate the coefficients $\alpha_4$ and $\alpha_5$, we obtain the quadratic relation of the energy spectrum,
  \begin{align}
   A\epsilon^2-2B\epsilon+C=0\, ,
  \end{align}
which is shown in \eqref{eq:eoe_disp} with the coefficients defined in \eqref{eq:ABC}.

\section{Computation of topological number of edge state}
\label{Comsgnh}
We compute the quantity $\text{sgn}(h_1)|_{k_i}$, which plays an essential role to determine the topological number associated with the edge states.
From Eqs. (\ref{alp1}) and (\ref{tilh4}) and the choice of the boundary conditions $\cos  \theta_1=\pi/2, a_2\ne 0, a_0=a_3=0$, we obtain
\begin{subequations}
\be
	\alpha_1&=&-a_1h_1\label{alpah1}
	\\
	&=&-a_1(2+\cos k_y+\cos k_z+ \gamma_1)\nonumber
	\\
	&=:&-a_1(\tilde{M}+ \gamma_1)\label{a1gam1}\,.
\ee
\end{subequations}
Eq.~(\ref{alpah1}) implies that $\text{sgn}(h_1)$ can be determined by $\text{sgn}(a_1\alpha_1)$. 
From Eqs. \eqref{eq:cosgam1&sinalp1}, we have 
\be\label{gam1alp1}
	\gamma_1^2=\alpha_1^2+1
	\, .
\ee
Hence, the combination of Eq (\ref{a1gam1}) and (\ref{gam1alp1}) provides a relation to determine the coefficient $\alpha_1$:
\begin{subequations}
\begin{align}
	(\alpha_1+a_1\tilde{M})^2 = a_1^2(\alpha_1^2+1)
	\ \iff \
	(1-a_1^2)\alpha_1^2-2a_1\tilde{M}\alpha_1+a_1^2(\tilde{M}^2-1) = 0 \,.
\end{align}
\end{subequations}
Since $1-a_1^2=a^2_2\ne0$, the discriminant of this quadratic equation is given by
\be
	\frac{1}{4} \Delta = a_1^2 \qty(\tilde{M}^2-1+\frac{1}{a_1^2})>0
	\, .
\ee
Therefore, there always exist two real roots which we call $\alpha_1^{(1)}$ and $\alpha_1^{(2)}$. 
In the calculation of the topological number associated with the edge states, we should take into account these two contributions.
We have the following relations for $\alpha_1^{(1)}$ and $\alpha_1^{(2)}$,
\begin{subequations}\label{eq:alp1&2}
\begin{align}
    \label{alp1+alp2}
	\alpha_1^{(1)}+\alpha_1^{(2)}=\frac{2a_1\tilde{M}}{1-a_1^2} & = 
	 \left\{ \begin{array}{ccl}
	    \displaystyle
         \frac{8a_1}{1-a_1^2} & \text{for} & k_y=k_z=0
         \\[1em] \displaystyle \frac{8a_1}{1-a_1^2}   & \text{for} & k_y=0, k_z=\pi/2  \text{ or } k_y=\pi/2, k_z=0
          \\[1em] \displaystyle
        0 & \text{for} &  k_y=k_z=\pi/2
            \end{array}\right.
    \\
    \label{alp1*alp2}
	\alpha_1^{(1)}\alpha_1^{(2)}=\frac{a_1^2(\tilde{M}^2-1)}{1-a_1^2} & = 
	 \left\{ \begin{array}{ccl}
	    \displaystyle
        \frac{15a_1^2}{1-a_1^2} >0& \text{for} & k_y=k_z=0
         \\[1em] \displaystyle \frac{3a_1^2}{1-a_1^2} >0   & \text{for} & k_y=0, k_z=\pi/2  \text{ or } k_y=\pi/2, k_z=0
          \\[1em] \displaystyle
        \frac{-a_1^2}{1-a_1^2} <0 & \text{for} &  k_y=k_z=\pi/2
                \end{array}\right.\,.
\end{align}
\end{subequations}
From these relations, we obtain 
\begin{align}
    \sum_{i=1,2}
	 \text{sgn}(a_1\alpha_1^{(i)}) = \left\{ \begin{array}{rcl}
        2& \text{for} & k_y=k_z=0  \text{ or } k_y=0, k_z=\pi/2  \text{ or } k_y=\pi/2, k_z=0
         \\  0   & \text{for} & k_y=k_z=\pi/2
                \end{array}\right.\,.
\end{align}
Hence, we conclude 
\be
    \sum_{i=1,2}
	 \left.\text{sgn}(h_1)\right|_{\alpha_1^{(i)}} = \left\{ \begin{array}{rcl}
        -2& \text{for} & k_y=k_z=0  \text{ or } k_y=0, k_z=\pi/2  \text{ or } k_y=\pi/2, k_z=0
         \\  0   & \text{for} & k_y=k_z=\pi/2
                \end{array}\right.\,,
\ee
which yields
\begin{align}
	\sum_{i=1,2}\left.\qty(\text{sgn}(h_1)\Big|_{(k_y,k_z)=(0,0)}+\text{sgn}(h_1)\Big|_{(k_y,k_z)=(\pi,\pi)}-\text{sgn}(h_1)\Big|_{(k_y,k_z)=(0,\pi)}-\text{sgn}(h_1)\Big|_{(k_y,k_z)=(\pi,0)})\right|_{\alpha_1^{(i)}}=2\,.
\end{align}


\bibliographystyle{utphys}
\bibliography{Latticeboundarycondition}

\end{document}